  \documentstyle[12pt]{article}
  
\newcommand{\be}{\begin{equation}}
\newcommand{\ee}{\end{equation}}
\newcommand{\bea}{\begin{eqnarray}}
\newcommand{\eea}{\end{eqnarray}}
\newcommand{\bref}[1]{(\ref{#1})}
\newcommand{\ep}{\epsilon} 
\newcommand{\T}{\theta}    \newcommand{\vp}{\varphi}
\newcommand{\A}{\alpha}    \newcommand{\B}{\beta}
\newcommand{\G}{\gamma}    
\newcommand{\D}{\delta}    \newcommand{\spc}{\varpi}

\newcommand{\m}{\mu}            
\newcommand{\r}{\rho}           
          
\newcommand{\w}{\omega}         \newcommand{\h}{\eta}

\newcommand{\dl}{\partial}   \newcommand{\nn}{\nonumber}

\def\ATs{anomalous transformations }
\def\NATs{non-anomalous transformations }

\newcommand{\sW}{{\scriptscriptstyle W}}
\newcommand{\sSW}{{\scriptscriptstyle SW}}

\def\pa{\partial} \def\da{\dagger} \def\dda{\ddagger}

\newcommand{\C}[1]{{\cal #1}}

\catcode`@=12
\title{{\bf  The WZ Term of  the Spinning String \\
\begin{center} and its On-shell Structure
\end{center}
}}
\author{{\sc J. Gomis},
        {\sc K.Kamimura$^\da$  and}
        {\sc R.Kuriki}$^{\dda}$\\
        \small{\it{Departament d'Estructura i Constituents
               de la Mat\`eria}}\\
        \small{\it{Universitat de Barcelona and }}\\
        \small{\it{Institut de F\'{\i}sica d'Altes Energies}}\\
        \small{\it{Diagonal, 647}}\\
        \small{\it{E-08028 BARCELONA}}\\
        \llap{$^\da$}%
        \small{\it{Department of Physics, Toho University}}\\
        \small{\it{Funabashi}}\\
        \small{\it{274 JAPAN}}\\
        \llap{$^\dda$}%
        \small{\it{Department of Physics, Tokyo Institute of Technology}}\\
        \small{\it{Oh-Okayama, Meguro}},
        \small{\it{Tokyo}}\\
        \small{\it{152 JAPAN}}\\
 \small{gomis@ecm.ub.es,
kamimura@jpnyitp.yukawa.kyoto-u.ac.jp,
r-kuriki@th.phys.titech.ac.jp}\\
}
\date{}

\begin{document}

\maketitle

\thispagestyle{empty}

\begin{abstract}
The Wess-Zumino term of the spinning string is constructed in terms
of their anomalies using an extended field-antifield formalism.
A new feature appears from a fact that the \NATs do not form a
sub-group. The algebra of the extended variables closes only using
the equations of motion derived from the WZ term.
\end{abstract}
\vskip 10mm

\vfill
\vbox{
%\draft
\hfill December 1995 \null\par
\hfill UB-ECM-PF 95/22\null\par
\hfill TOHO-FP-9553\null\par
\hfill }\null

\clearpage

%%%%%%%%%%%%%%%%%%%%%%%%%%%%%%%%%%%%%%%%%%%%%%%%%%%%%%%%%%%%%%%
\baselineskip 6.5mm

\section{Introduction}
\indent

In a recent paper\cite{gkpz} we have analyzed the form of the
Wess-Zumino (WZ) term
at one loop for general anomalous on-shell gauge theories using an
extended field-antifield formalism. The gauge degrees of freedom that
become propagating at quantum level are introduced as dynamical fields.
The antifield independent part of the WZ term, in the classical basis,
has the usual form \cite{z}\cite{gp1}\cite{gp2} and is expressed
in terms of the anomalies and the finite gauge transformations.
The full WZ term is obtained by the cohomological reconstruction
procedure \cite{h} \cite{vp}.

In this paper we analyze the system of spinning string \cite{dz}\cite{bdh}
as an  example of anomalous gauge theories with open algebras.
The Weyl and super-Weyl transformations become anomalous in
regularizations in which the diffeomorphism invariance is preserved.
The Liouville and the super-Liouville fields are introduced
as extra degrees of freedom in a natural way and are propagating
at quantum level. The WZ term is constructed from expressions of
the anomaly and the finite anomalous transformations.

The on-shell structure of the extended formalism has a new phenomena
that the algebra of two SUSY transformations of the super Liouville
field does not close at classical level.
Therefore we cannot obtain a solution of the classical master equation
in the extended formalism at this level.
However the non-closure term vanishes on-shell of its equations of motion
obtained from the WZ term describing dynamical properties of the
super Liouville fields.
When we consider the quantum action including the WZ term and perform a
canonical scale transformation depending on $\hbar$ we can
isolate  terms independent  of $\hbar$ as the solution of
the classical master equation.
The residual term is a quantum correction and is proportional to
$\sqrt\hbar$.
It is the background term and cancels the anomaly at one loop.

This paper is organized as follows.
In section 2, we briefly outline the proper solution of the classical
master equation in the space of local functional of fields and antifields.
In section 3 the WZ action is constructed from the super-Weyl
anomaly using the form of finite transformations.
Section 4 is devoted to discussion of the extended field-antifield
formalism for the spinning string.
Summary and discussion are given in Section 5.

%%%%%%%%%%%%%%%%%%%%%%%%%%%%%%%%%%%%%%%%%%%%%%%%%%%%%%%%%%%%%%%%

\section{Classical symmetry and proper solution of classical master
equation}
\indent

In this section we will review the classical gauge symmetries of the
spinning string and the construction of solution of classical
master equation.

The classical action of spinning string is given by
\cite{dz}\cite{bdh}\footnote{We follow the notation of \cite{GSW}.}
\bea
S_0~=~-\int d^2x~e~\biggl[~\frac{1}{2}
\bigl(g^{\A\B}\dl_\A X\dl_\B X-i\overline\psi\r^\A\nabla_\A\psi\bigr)
%\nn\\~~~~~~~~~~~~~~~~~~
+\overline\chi_\A\r^\B\r^\A\psi\dl_\B X
+\frac{1}{4}(\overline\psi\psi)(\overline\chi_\A\r^\B\r^\A
\chi_\B)~\biggr],
\label{str}
\eea
where $X^\m$ and $\psi^\m$ ( $\m$=0,...,$D$-1 ) are, respectively,
the bosonic and fermionic string variables and we suppress the
space-time indices $\mu$. The zwei-bein field and gravitino are denoted by
$e_\A{}^a$ and $\chi_\A$, respectively.
The covariant derivative $\nabla_\A$ is defined by
\bea
\nabla_\alpha\equiv\partial_\alpha+\frac{1}{2}\spc_\alpha\rho_5,
\eea
where $\spc_\A$ is the spin connection defined by\footnote
{~$\ep^{01}=-\ep_{01}=1$ both for the world and local
tensors $\ep^{\A\B}$ and $\ep^{ab}$.}
\bea
\spc_\A~=~\spc_\A^{(0)}~+~~\spc_\A^{(1)}~;~~~~~
\spc_\A^{(0)}~=~\frac{1}{e}e_\A^{~a}\ep^{\B \G}\pa_\B e_{\G a},~~~~
\spc_\A^{(1)}~=~2 i \bar{\chi}_\A \r_5\r^\B \chi_\B.
\eea

The action is invariant under the world-sheet reparametrization,
local Lorentz rotation, local supersymmetry(SUSY), Weyl
and super-Weyl transformations.
%\bea
%\label{infgauge}
%\eea
Since these transformations are independent the algebra is irreducible.
In contrast with the bosonic string the algebra is open since the
commutator of two SUSY transformations of the fermion $\psi$ closes
only on-shell of the classical equation of
motion $S_{0,\psi}\approx 0$. That is
\bea
\left[ \delta_{SUSY},\delta'_{SUSY}\right]\psi~=~
\delta''\psi~
-\frac{1}{e}\epsilon{'}\left(S_{0,\psi}\epsilon\right)
+\frac{1}{e}\epsilon\left(S_{0,\psi}\epsilon{'}\right),
\label{infalg}
\eea
where
$\epsilon$ and $\epsilon{'}$ are the parameters of the local
super-symmetry transformations.
\bea
S_{0,\psi}~ \equiv~ \frac{\delta^r{S_0}}{\delta\psi}~
=~e~\overline{\left\{i\rho^\alpha\nabla_\alpha\psi-
\rho^\alpha\rho^\beta\chi_\alpha
\left(\partial_\beta{X}-\bar{\psi}\chi_\beta\right)\right\}}
\eea
are the equations of motion of the fermionic matter.
The first term of r.h.s. of \bref{infalg},
$\delta''\psi$, represents a sum of
reparametrization, local Lorentz and local super-symmetry
transformations.

The classical algebraic structures are nicely formulated as the classical
master equation ( CME ) in the field antifield formalism
\cite{zin}\cite{bv}\footnote{For reviews see \cite{ht}\cite{gps}\cite{tp}}.
It is expressed in terms of anti-bracket as
\be~(S,S)~=~0.~
\label{CME}
\ee

The solution of the CME for the spinning string is \cite{Go3}.
\bea
S~&&=~S_0~
\nn\\&&
+~X^*~(\partial_\A X~C^\A+\overline\psi\w~) \nn \\
&& +~\psi^*~(\partial_\A\psi~C^\A~+~\frac{1}{2}\r_5\psi~C_L
-\frac{1}{4}\psi~C_\sW-i\r^\A\w(\partial_\A X-\overline\psi~\chi_\A))
\nn \\
&& +~{e^\A_{~a}}^*~(\partial_\B e_\A{}^a~C^\B
+\ep^a{}_b~e_\A{}^b~C_L+\frac{1}{2}~e_\A{}^a~C_\sW
+2i\overline{\chi_\A}~\r^a~\w~)
\nn \\
&& +~{\chi^\A}^*~(~\partial_\B\chi_\A~C^\B
+\chi_\B~\partial_\A C^\B
+\frac{1}{2}\r_5\chi_\A~C_L~+~\frac{1}{4}\chi_\A~C_\sW
+\frac{i}{4}\r_\A~\h_{\scriptscriptstyle W}+
\nabla_\A\w~) \nn \\
&& +~{C_\A}^*~(~\partial_\B C^\A~C^\B
-i\overline\w\r^\A\w ~) \nn \\
&& +~ {C_L}^*~(~\dl_\A C_L~C^\A
+\frac{1}{2}\overline\w\r_5\h_{\scriptscriptstyle W}
-  \spc_\A~i\overline\w\r^\A\w) \nn \\
&& +~{\w}^*~(~\partial_\A\w~C^\A~+~\frac{1}{2}\r_5\w~C_L~+~
\frac{1}{4}\w~ C_\sW ~+~\chi_\A~(i\overline\w\r^\A\w)) \nn \\
&& +~{C_\sW}^*~(~\partial_\A C_\sW~C^\A
-\overline\w\h_{\scriptscriptstyle W} ~) \nn \\
&&+~{\h_{\scriptscriptstyle W}}^*~(
\partial_\A \h_{\scriptscriptstyle W}~C^\A
+\frac{1}{2}\r_5\h_{\scriptscriptstyle W}~C_L
-\frac{1}{4}\h_{\scriptscriptstyle W}~C_\sW
-i\r^\A\w(\partial_\A C_\sW~+~\overline\chi_\A
\h_{\scriptscriptstyle W})
-\frac{4}{e} \w ~(\overline\w \r_5 T)~) \nn\\
&&-~\frac{1}{2 e}(\psi^*\w)(\psi^*\w),
\label{solCME}
\eea
where $~C^\alpha$, $\omega$, $C_L$, $C_\sW$ and $\eta_\sW$ are ghosts for
reparametrization, local supersymmetry, local Lorentz rotation,
Weyl and  super-Weyl transformations respectively.
$T$ is defined as $~T\equiv\ep^{\A\B}\nabla_\A\chi_\B$.

The BRST transformation in the space of fields and
antifields is generated by $S$,
\be \delta ~\cdot~=~(~\cdot~,~S).
\ee
It is nilpotent {\it off-shell} as the result of the CME \bref{CME}.
Note that the solution  $S$ in  \bref{solCME} contains a quadratic
term of anti-field $\psi^*$ and the BRST transformation of the fermionic
matter $\psi$ depends on the antifield reflecting the properties of the
on-shell algebra \bref{infalg}.

It is also useful to consider the following BRST operator in the
classical basis. It acts on functional depending only on the fields as
\be
 \delta_0~ \cdot~  =~(~\cdot~, {S})|_{\Phi^*=0}.
\ee
It is nilpotent {\it on-shell} of the classical equations of motion of the
fermionic matter $\psi$,
\be
\delta_0^2~\psi~
={\frac{ \omega}{ e}}
({S_{0,\psi}}\omega).
\label{e0}
\ee

%%%%%%%%%%%%%%%%%%%%%%%%%%%%%%%%%%%%%%%%%%%%%%%%%%%%%%%%%%%%%%%%%%%%%
\section{ The Super Liouville action}
\indent

Here we will construct the WZ term for the spinning string in terms
of the anomalies of the theory. The antifield independent part of
the anomaly in the diffeomorphism invariant regularizations
is known as the super-Weyl anomaly. It has been given explicitly
in a Hamiltonian BRST procedure \cite{Fuji2}
\bea
{\cal A}~=~-i\hbar k \int d^2x[~ 2~(\ep^{\A\B}\pa_\A \spc_\B)~C_\sW~-~
4~{\bar T}~\r_5~\h_\sW~]~\equiv~\, {\C{A}}_{a}(\phi) c^a,
\label{anomaly}
\eea
where $~k~$ is a constant proportional to $(10-D)$ and
$~c^a~$ denote ghosts for the anomalous transformations;
{\it i.e.} the Weyl and super-Weyl ghosts, $C_\sW$ and $\h_\sW$.
%index $~a~$ runs over that of the anomalous transformations;
%{\it i.e.} the Weyl and super-Weyl ones.
The super-Weyl anomaly \bref{anomaly} verifies
\bea
\delta_0{\cal A}~=~0.
\label{del0a}
\eea
The cohomological reconstruction procedure \cite{h}\cite{vp}
allows us to reconstruct a BRST invariant object depending on fields and
antifields from a weakly $\delta_0$ invariant antifield independent object.
Since \bref{del0a} holds off-shell we can conclude that  $\C{A}$ is
actually  the complete anomaly and does not depend on the antifields.
This fact can be understood intuitively by taking into account the fact
that the algebra is open only for the fermionic matter $\psi$
while the super-Weyl anomaly ${\cal A}$ does not contain $\psi$.
The corresponding WZ term $~\C{M}_{1}(\phi,\T^a)$
will depend only on the fields and verifies
\be
\label{WZeq AI C1}
\delta_0  \C{M}_{1}(\phi,\T^a) ~=~  i\, {\C{A}}_{a}(\phi) c^a \,,
\ee
where $\T^a$ indicates parameters of the anomalous subgroup.
This antifield independent part of the WZ term becomes actually
the full WZ term satisfying
\be
\label{WZeq AI C2}
\delta\C{M}_{1}(\phi,\T^a) ~=~  i\, {\C{A}}_{a}(\phi) c^a
\ee
in this case.

The WZ term can be obtained in terms of the anomaly as
\bea
\label{WZ2}
{\cal M}_{1}\left(\phi,\T\right) = \, -i\int_0^1{d t}\,{\cal A}_a
\left({F}\left(\phi,t\T\right)\right)
\lambda^a_{~b}\left(t\T,\phi\right)\T^b.
\eea
Here $ F^i(\phi,\T^a)$ are the finite anomalous transformations
of $\phi^i$ and satisfy the on-shell composition law
\be
F^i (F (\phi,\theta),\theta')=
F^i (\phi,\varphi(\theta,\theta';\phi)) +
M^{ij}(\theta,\theta';\phi) {\cal S}_{0,j}(\phi),
\label{openF}
\ee
where $ M^{ij}(\theta,\theta';\phi)$ represents the open structure at finite
level \cite{gkpz}.
$\lambda^a_{~b}$ is an inverse of   $\mu^b_{~a}$
defined by using the composition function of the quasi-group as;
\be
\mu^a_{~b}(\T,\phi)= \left.\frac{\partial
\vp^a(\T,\T',\phi)}{\partial\T'^b}\right|_{ \T'=0}.
\label{mu}
\ee

For the spinning string these transformations are
\bea
\tilde X^\mu~=~X^\mu,~
\tilde\psi^\mu~=~e^{\frac{\sigma}{4}}\psi^\mu,~
\tilde e_\A^{~a}~=~e^{\frac{\sigma}{2}}e_\A^{~a},~
\tilde \chi_\A~=~e^{\frac{\sigma}{4}}(\chi_\A+\frac{i}{4}\rho_\A\eta),
\label{finite}
\eea
where $\sigma$ and $\eta$ are, respectively, the Weyl and the super Weyl
finite transformation parameters.
The composition law of two finite
anomalous transformations is expressed by
\bea
\varphi^\sigma\left(\theta,\theta{'}\right)=\sigma+\sigma{'},~~~~~~
\varphi^\eta\left(\theta,\theta{'}\right)=\eta+e^{\frac{\sigma}{4}}
\eta{'}
\label{compo}
\eea
and the $\lambda^a_{~b}$ is given by $\pmatrix{1&0\cr 0&e^{-\frac\sigma 4}}$.

The WZ term is constructed using \bref{WZ2} as,
\bea
{\cal M}_1~=~
\hbar k S^{SL},
\eea
where
\bea
&&S^{SL}~\equiv~ S^{SL}_0~+~S^{SL}_1, \\
\label{sla}
&&S^{SL}_0=- ~\int d^2x ~e~\Bigl\{\frac{1}{2}
\bigl(g^{\A\B}\dl_\A \sigma\dl_\B \sigma
-i\overline\h\r^\A\nabla_\A\h\bigr)
-\overline\chi_\A\r^\B\r^\A\h\dl_\B \sigma
+\frac{1}{4}\overline\h\h\overline\chi_\A\r^\B\r^\A
\chi_\B \Bigr\},  \\
&&S^{SL}_1=-~\int d^2x
[2 \ep^{\A\B}\pa_\A\spc_\B~\sigma~-~4~{\overline T}\r_5 \h].
\eea
It is the super-Liouville action found in \cite{p}\cite{Fuji1}.
Remember $S^{SL}_0$ is obtained from the classical action by the
replacements $X \rightarrow  \sigma$ and $\psi \rightarrow - \h$
and $S^{SL}_1$ is obtained from the anomaly by the replacements
$C_\sW \rightarrow \sigma$ and $\h_\sW \rightarrow \h$.

The transformation properties of the extra variables are determined
by requirements that $F^i(\phi^j,\T^a)$'s in \bref{finite} are
invariant under the \ATs and covariant under \NATs.
They are
\bea
&&\delta_0\sigma~=~\partial_\A\sigma~\ep^\A-\overline\h\ep -\ep_\sW~\nn\\
&&\delta_0\h  ~=~\partial_\A\h \ep^\A+\frac{1}{2}\r_5\h \ep_L
+i\r^\A\ep(\partial_\A\sigma+{\overline\h}\chi_\A)-\frac14 \h \ep_\sW-
\ep_{\sSW}.
\label{deltheta}
\eea
The super-Liouville action is invariant under \NATs;
reparametrization ($\ep^\A$),
local Lorentz ($\ep_L$) and SUSY transformation ($\ep$).
Under \ATs; Weyl ($\ep_\sW$) and super Weyl ($\ep_{\sSW}$),
$S^{SL}$ is not invariant.
It is the source of anomaly cancellation.

%%%%%%%%%%%%%%%%%%%%%%%%%%%%%%%%%%%%%%%%%%%%%%%%%%%%%%%%%%%%%%%%%%
\section {On-shell structure of the extended formalism}
\indent

Now we want to study the BRST transformations of the extra variables
in more detail. Let us now construct the  extended action in the space of
fields and antifields.
We start  considering a classical action that includes the
transformation properties of the Liouville and super-Liouville fields
as well as the original ones\cite{gp1}\cite{gp2},
\bea
S^{(ext)}~=&&~S~+~\sigma^*
(\partial_\A\sigma~C^\A-\overline\h\w -C_\sW)~\nn\\
&&+~\h^* (\partial_\A\h C^\A
+\frac{1}{2}\r_5\h C_L
-\frac{1}{4}\h C_\sW
+i\r^\A\w(\partial_\A\sigma+{\overline\h}\chi_\A)
-\h_{\scriptscriptstyle W}).
\label{Sext}
\eea
$S^{(ext)}$ does not verify the classical master equation
because the algebra of the transformations of the
extra variables \bref{deltheta} is not closed, or equivalently its BRST
transformations do not satisfy the nilpotency,
\be
\delta_0^2~\eta~=~-(\bar\omega a)~\omega,~~~~~
a~\equiv~i\rho^\A\nabla_\A\h+\rho^\A\rho^\B
\chi_\A(\pa_\B\sigma+\bar\eta\chi_\B)-\frac{4}{e}\rho_5T.
\label{e1}
\ee
Note $a$ is the Euler derivative of the super Liouville action
with respect to $\eta$.

The situation is similar to the transformation of the
fermionic matter $\psi$ in which the closure of the algebra requires
the classical equation of motion of $\psi$.
However we don't have the classical equation of motion
for the extra variable $\eta$.
The infinitesimal algebra closes and has nilpotent BRST transformation
when we take the WZ term as is describing
the dynamical action of the super Liouville fields.
In fact the WZ term enters in the quantum action.
It has the coefficient proportional to $\hbar$
while $\T^*\D_0 \T$ term does not contain $\hbar$.
In addition bilinear terms of antifields reflecting the open algebra
structure of the extra variables, symbolically
expressed as  ${(\T^*)}^2$, are proportional to $\hbar^{-1}$,
\be
W~=~S~+\hbar S^{WZ}~+~\T^*~\D_0 \T~+~\frac{1}{\hbar}{(\T^*)}^2+...
\ee
In case of absence of ${(\T^*)}^2$ term we can take the classical limit
$\hbar \rightarrow 0$ to find $S^{(ext)}$. In presence of ${(\T^*)}^2$
term we will take the classical limit only after making a canonical
transformation \cite{jst} \cite{vp} \cite{gp2}
\be
\T~=~{\hbar}^{-1/2}\tilde\T,~~~~~
\T^*~=~{\hbar}^{1/2}\tilde\T^*.
\label{ct}
\ee
After this transformation
%(in situations in which there remains
%no negative power term of $\hbar $, as in the spinning string case)
\be
W~=~W_0~+~\hbar^{1/2}~M_{1/2}+....
\ee
The classical limit gives $W_0$ and is the extended proper solution
of CME.
The second term $M_{{1}/{2}}$ is the first order quantum correction
referred as the background term \cite{gp2}.

%%%%%%%%%%%%%%%%%%%%%%%%%%%%%%%%%%%%%%%%%%%%%%%%%%%%%%%%%%
In the present model $W$ is
\bea
W~=~S~+~\hbar k S^{SL}~+~\sigma^* \D_0\sigma ~+~\h^* \D_0 \h -
\frac{1}{2 \hbar k e}(\h^*\omega)(\h^*\omega).
\eea
It satisfies
\bea
(W,W)~=~2i~{\cal A}.
\eea
After the  canonical transformation \bref{ct}
we have that $W_0$ and $M_{{1}/{2}}$ are given by
\bea W_0~&&=~S~+~k S_0^{SL}~
+~\sigma^*(\partial_\A\sigma~C^\A-\overline\h\w )~\nn\\
&&+~ \h^*(\partial_\A\h C^\A+\frac{1}{2}\r_5\h C_L
-\frac{1}{4}\h C_\sW
+i\r^\A\w(\partial_\A\sigma+{\overline\h}\chi_\A))
-\frac{1}{2ke}(\h^* \w)^2,
\label{W0}\\
M_{1/2}~&&=~-k[2 \ep^{\A\B}\pa_\A\spc_\B~\sigma~-~4~{\overline T}\r_5 \h]
-\sigma^*C_\sW-\h^*\h_\sW.
\eea
$\sigma$ and $\h$ in the above expressions are understood to be new
variables $\tilde\sigma$ and $\tilde\h$.
They satisfy
\be
(W_0,W_0)~=~0,~~~~(W_0,M_{1/2})~=~0,~
\ee
\be~~~
\hbar(M_{1/2},M_{1/2})~=~2\hbar k
[2 \ep^{\A\B}\pa_\A\spc_\B~C_\sW~-~4~{\overline T}\r_5 \h_\sW]~=~2i{\cal A}.
\label{MM}
\ee
$W_0$ should be regarded as an extended classical action.
The first two equations mean that
both $W_0$ and $M_{1/2}$ are invariant under BRST transformation
generated by $W_0$.
The last expression has the form of the super-Weyl anomaly.

In  $W_0$ the extra variables  $\sigma$ and $\h$ are coupled
 with super-gravity multiplet
exactly in the same way as $X^\mu$ and $\psi^\mu$.
$W_0$ is expressed in the same form as
$S$ in \bref{CME} if $(\sigma,\h)$ is regarded as (D+1)-th component of
$(X^\mu,\psi^\mu)$. The Liouville and super-Liouville fields give
additional contribution  to the anomaly and shift the proportional constant
$(10-D)$ in the coefficient $k$ of the anomaly to $~(10-(D+1))~=~(9-D)$.
After this replacement the background term will cancel the
anomaly and the quantum master equation is satisfied at one-loop.
%%%%%%%%%%%%%%%%%%%%%%%%%%%%%%%%%%%%%%%%%%%%%%%%%%%%%%%%%%%

\section{Summary and discussion}
\indent

We have constructed the extended
field-antifield formalism of the
spinning string as an example of an anomalous theory with a
 open and irreducible gauge algebra. We have followed the general
discussion and constructed the WZ term based on the
super-Weyl anomaly via the finite transformations\cite{gkpz}.

A new phenomena appears  at classical
level since there is no solution of the classical master equation in the
extended space. The nilpotency of BRST transformation for the
fermionic Liouville field $\eta$ is not satisfied.
The non-vanishing term is actually proportional to the dynamical
equation of motion for the fermionic Liouville field obtained from
the WZ term, namely super-Liouville action.
The extended solution including the super-Liouville action
is not regarded as a classical object. In this situation quadratic
terms of antifield of fermionic Liouville field appears in the
quantum action and are proportional to $\hbar^{-1}$. In order to
take the classical limit $\hbar\rightarrow{0}$, one need to perform
the canonical transformation \bref{ct}.
After the canonical transformation is performed  we can split
the extended action into an action $W_0$ verifying the classical master
equation  and the background term $M_{1/2}$ which is proportional to
$\hbar^{\frac{1}{2}}$.
$W_0$ defines an off-shell nilpotent BRST transformation and
the background term cancels the anomaly at one-loop level.

The origin of the non-nilpotency of the extra variables is the
fact that commutator of two non-anomalous (SUSY) transformations
do not close but gives  anomalous (super-Weyl) transformations.
This phenomena generally occurs when a part of gauge symmetries
is broken due to anomalies but the non-anomalous transformations
do not form a sub-algebra.
The analysis of this phenomena in a generic gauge theory will be
discussed elsewhere \cite{gkpz2}.

%%%%%%%%%%%%%%%%%%%%%%%%%%%%%%%%%%%%%%%%%%%%%%%%%%%%%%%%%%%

\vskip 6mm

{\bf Acknowledgments}

 This work has
been partially supported by CYCYT under contract number AEN93-0695,  by
Comissionat per Universitats i Recerca de la Generalitat de Catalunya and
by Commission of European Communities contract CHRX-CT93-0362(04).
One of the author (KK) would thank "{\bf Nukada Grant}" for financial
support.

\end{document}